\documentclass[aps,prx,twocolumn,superscriptaddress,reprint,eprint,longbibliography]{revtex4-2}

\usepackage[margin=1in]{geometry}
\usepackage{amsmath,amssymb,amsthm}
\usepackage{enumitem}
\usepackage{graphicx}
\usepackage[normalem]{ulem}

\usepackage[colorlinks=true,
            linkcolor=blue,
            citecolor=blue,
            urlcolor=blue]{hyperref}
\usepackage{xcolor}
\usepackage{braket} %
\usepackage{comment} %
\usepackage{array}
\usepackage{subcaption} 
\usepackage{caption}
\newtheorem{theorem}{Theorem}

\newcommand{\code}[1]{\texttt{#1}}

\begin{document}
\title{Testing Quantum Advantage in High-Frequency Trading with Market-Derived Nonlocal Games}
\author{Daiwei Zhu}
\thanks{These authors contributed equally.}
\affiliation{IonQ Inc., College Park, MD 20740, USA}
\author{Brendan Pankovich}
\thanks{These authors contributed equally.}
\affiliation{IonQ Inc., College Park, MD 20740, USA}
\author{Alireza Marefat Khah}
\thanks{These authors contributed equally.}
\affiliation{IonQ Inc., College Park, MD 20740, USA}
\author{Haakam Aujla}
\affiliation{IonQ Inc., College Park, MD 20740, USA}
\author{Ramis Movassagh}
\affiliation{IonQ Inc., College Park, MD 20740, USA}
\author{Martin Roetteler}
\affiliation{IonQ Inc., College Park, MD 20740, USA}
\author{Johannes Borregaard}
\affiliation{IonQ Inc., College Park, MD 20740, USA}
\date{\today }

\begin{abstract}
Non-classical correlations enabled by entanglement could be exploited for distributed coordination tasks with communication constraints such as high-frequency trading in fragmented markets. However, previous studies have largely specified the game and its payoff structure theoretically, using toy market models without testing how well these models reflect observed market behavior. Here, we introduce market-derived nonlocal games, in which historical order-book data and backtests of joint actions determine the game’s inputs and payoffs, rather than a prespecified idealized model. We start from classical trading strategies based on market considerations, using quantum correlations only to coordinate the agents’ decisions. We compute exact classical optima and quantum upper bounds for the payoff. For games with a possible quantum advantage, we introduce an optimization algorithm to find explicit entangled strategies and end-to-end backtest them on historical data. We demonstrate the approach in a two-agent, three-stock cross-venue arbitrage setting. The game derived from the observed market data shows no resolved quantum-classical gap at the measured interaction scales. 
Controlled variations in liquidity competition and cross-stock repricing then show how the approach identifies regimes supporting quantum advantage, constructs explicit strategies, and tests whether their advantage survives end-to-end backtesting on modified historical market data.
\end{abstract}

\maketitle

\section{Introduction}
\label{sec:intro}

Quantum nonlocality is a key distinguishing feature that separates classical and quantum information science. The nonlocality of entangled quantum systems continues to attract significant interest from both foundational and applied perspectives, underpinning a broad range of concepts including device-independent quantum key distribution~\cite{acin2006,vazirani2014,zhang2022,nadlinger2022}, quantum self-testing~\cite{McKague2012,Zhang2018}, and foundational tests of quantumness~\cite{Bell1964,hensen2015,Borregaard2025,covey2025,fromonteil2025}. 
More recently, quantum nonlocal games have been explored as a means for non-communicating parties to outperform classical strategies in distributed coordination tasks with communication constraints~\cite{brassard2005,ding2026}. The proposed applications include high-frequency trading in fragmented markets~\cite{blm2015,ding2024coordinating} and load balancing in distributed computing~\cite{hasanpour2017,arun2025,dasilva2026}. 
These proposals have remained largely conceptual, mapping idealized nonlocal games with known quantum-classical separations onto toy market models, leaving it unclear how well those models capture real market behavior.

To address this gap, we introduce market-derived nonlocal games, in which the game is derived from observed market dynamics rather than selected in advance.
We start from a classical trading strategy that specifies each agent's trading signals and available actions based on market dynamics and financial objectives. Quantum correlations are introduced only to coordinate the agents' choices among those actions. 
Backtesting on historical market data determines the joint distribution of agents' local trading signals and the payoffs associated with their joint actions. These quantities define a nonlocal game in which agents choose their actions without communicating during a round. For each market-derived game, we compute the exact classical optimum  by exhaustively enumerating all combinations of the agents’ deterministic strategies. We compare this value with quantum upper bounds from the Navascu\'es--Pironio--Ac\'in (NPA) hierarchy~\cite{Navascues2008}. Where an upper bound leaves room for a quantum-classical gap, we introduce an optimization algorithm over the shared state and local measurements of agents to search for explicit entangled strategies. Their achieved payoff shows whether the gap can be realized.

We use a two-agent, three-stock cross-venue arbitrage setting to validate the market-derived-game approach on historical market data using the \texttt{NautilusTrader} backtesting engine~\cite{nautilustrader}. From these data, we identify and estimate the payoff effects of competition for shared liquidity and cross-stock repricing. At their measured scales, the resulting game shows no resolved quantum-classical gap. We then strengthen these empirically identified interactions in controlled benchmarks to determine when a gap emerges and whether explicit strategies found in those regimes retain an advantage in end-to-end backtests.

\begin{figure*}[t]
    \centering
    \includegraphics[width=\textwidth]{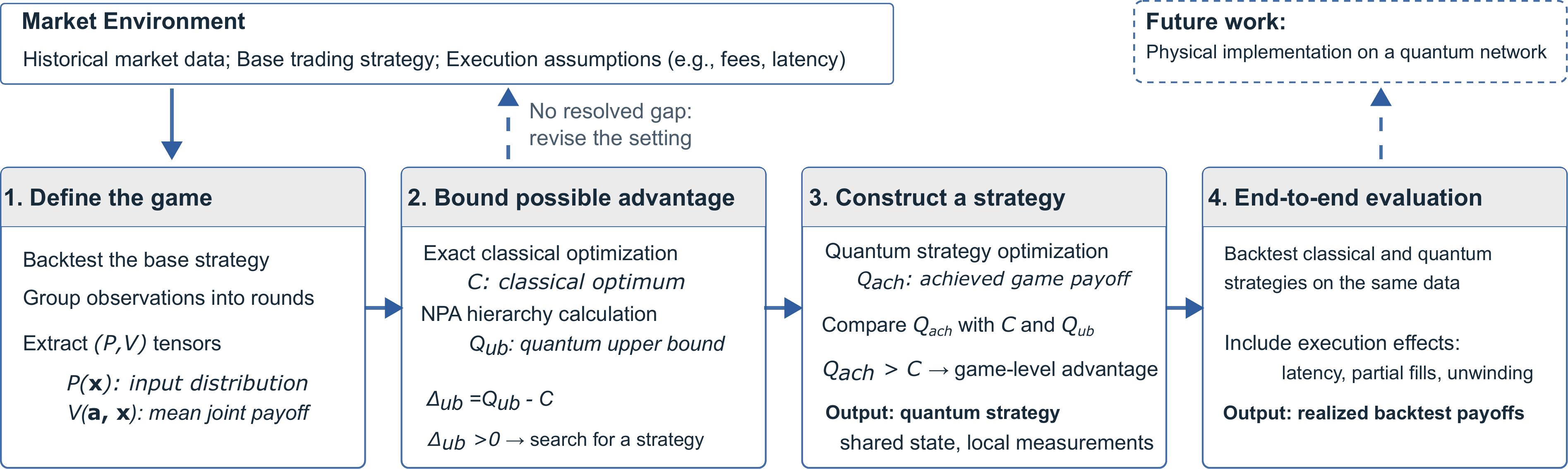}
    \caption{Testing Quantum Advantage in High-Frequency Trading with Market-Derived Nonlocal Games. (1) The environment is defined by historical market data, a base trading strategy specifying local signals and available actions, and execution assumptions. Backtests yield the input distribution $P$ and the mean joint-payoff tensor $V$. The market-derived nonlocal game is fully specified by the $P,V$ tensors. (2) Exact classical optimization gives $C$, while the NPA hierarchy gives an upper bound $Q_{\mathrm{ub}}$ on the optimal quantum value $Q$. The upper-bound gap $\Delta_{\mathrm{ub}}=Q_{\mathrm{ub}}-C$ identifies scope for an advantage but does not establish its attainability. Settings with no numerically resolved gap are reconsidered. (3) Quantum strategy optimization constructs a shared state and local measurements with achieved payoff $Q_{\mathrm{ach}}\leq Q\leq Q_{\mathrm{ub}}$; $Q_{\mathrm{ach}}>C$ establishes an advantage in the game. (4) Classical and quantum strategies are evaluated on the same market data under the same execution assumptions, where we compare the realized backtest payoffs using simulated quantum correlations. Physical implementation on a quantum network is future work.}
    \label{fig:workflow}
\end{figure*}

Our benchmark shows that the approach can derive a game from historical market data, identify the market interactions that shape its payoffs, and test explicit strategies in controlled regimes where a quantum-classical gap emerges. This cross-venue example is only one of many possible applications. Each financially grounded trading strategy and market environment defines a new market-derived nonlocal game. Other classical HFT strategies, including prediction-based strategies~\cite{kearns2013machine} and strategies that manage inventory risk or trade off expected profit against risk measures such as profit-and-loss (PnL) variance~\cite{cartea2015risk,fodra2012}, could define further market-derived games in which one can search for quantum advantages. The cost of computing classical optima grows with the number of agents, signals, and actions, while tighter NPA bounds require higher hierarchy levels~\cite{Navascues2008}. However, the computational size of a game can be controlled in part by restricting the agents’ local action sets to choices relevant to the underlying trading strategy, while retaining the market interactions of interest. The binary decision to trade or not for a given stock signal in our example is one such restriction. This opens a systematic research direction for identifying  realistic HFT strategies and market conditions amenable to practical quantum advantage.

\section{From market data to nonlocal games}
\label{sec:framework-overview}

Testing quantum advantage in a trading environment requires both identifying advantageous coordination strategies and determining whether their gains survive market execution. We therefore connect market data to a nonlocal game, bound the possible advantage,  explicitly construct quantum strategies, and backtest them end-to-end~(Fig.~\ref{fig:workflow}).

\subsection{Defining the nonlocal game}
\label{sec:nonlocal-game}

The first stage extracts a multiplayer nonlocal game~\cite{brassard2005,kempe2007} directly from a real market environment using historical order books, a base trading strategy, and market parameters such as fees and communication latencies. 

High-frequency trading refers to automated trading strategies that react to market data on timescales of microseconds to milliseconds.
Trading occurs on electronic exchanges, or \emph{venues}, each of which maintains an order book of standing buy (\emph{bid}) and sell (\emph{ask}) orders and executes incoming orders according to price and arrival time.
Venues publish updates to their books and the same asset is often traded on several venues where quoted prices can briefly differ.

An $n$-agent market-derived nonlocal game is constructed as follows.
The $i$th agent receives a local trading signal, encoded as input $x_i$, and returns an output $a_i$, which specifies its trading action.
Let $\vec{x} = (x_1, \ldots,x_n)$ and $\vec{a} = (a_1,\ldots,a_n)$ organize the collective set of inputs and outputs across all $n$ agents.
The trading signal and the allowed actions are set by a financially grounded base trading strategy.
For example, in a simple arbitrage strategy, an agent may observe that the ask price at one venue is below the bid price at another and respond by submitting buy and sell orders at the venues.
The subsequent payoff depends on whether those orders execute and the resulting spread after costs~\footnote{In the United States the venues are subject to regulation. The best bid and ask across all venues form what is known as the national best bid and offer (NBBO), and, for instance, a venue may not execute an order at a price worse than one displayed at another venue, nor display a quote that equals or crosses another venue's quote on the opposite side. Nonetheless, quotes can briefly appear crossed in the venues' direct feeds due to propagation delays between venues. Such short-lived dislocations motivate the cross-venue arbitrage example.}.

Once the base strategy is fixed, we run it on historical market data using a backtesting engine~\cite{nautilustrader} with the agents acting independently, i.e.,\ without coordinating, and solely based on their local view of the market.
We divide the historical timeline into non-overlapping windows of width $w$, with boundaries agreed upon in advance.
Each nonempty window defines one round of the nonlocal game. 
Within each window, we record the market signals observed by each agent and map them to finite input alphabets.
For example, an input can indicate which asset, if any, carries a trading signal of the base strategy within the window. For the arbitrage strategy, this is an asset whose ask price at one venue lies below its bid price at another.
This gives a sequence of input tuples $\{\vec{x}(\tau)\}_{\tau=0}^{N_w-1}$, where $N_w$ is the number of nonempty windows. 

We discard windows in which no agent observes a trading signal, $\vec{x}=\vec{0}$, and normalize the input distribution $P(\vec{x})$ over the remaining windows. Expected game payoffs are therefore expressed per nonempty round and session payoff predictions are obtained by multiplying by the number of nonempty rounds.
To determine the payoff tensor $V(\vec{a},\vec{x})$, we replay the historical record under each allowed joint action tuple and average the realized payoff over the occurrences of the corresponding input.
Since the input tuples are coarse-grained representations of the market, the same input and action tuple can yield different payoffs on different occasions. The tensor therefore records the mean payoff over the trading period for each combination of inputs and actions.

The resulting pair $(P,V)$ defines a multiplayer nonlocal game derived from the measured market environment.
We consider real-valued payoffs, including losses, rather than restricting $V$ to the binary win-or-lose condition that is often used~\cite{cleve2010}. 
As both tensors are derived from backtesting on historical data, rather than assumed, the resulting game inherits the statistical structure of the real environment. Hence any exhibited classical--quantum gap is grounded in observed dynamics.

\subsection{Classical optimum and quantum bound}
\label{sec:classical-quantum-bounds}

In the second stage we determine if the market-derived nonlocal game can support a quantum advantage. 
To this end, we compute the classical optimum and compare it with an upper bound on the quantum value.

A strategy induces a conditional output probability distribution, or \emph{behavior}, $p(\vec{a}\mid \vec{x})$. 
Its expected payoff is
\begin{equation}
  \Pi[p]=\sum_{\vec{a},\vec{x}} P(\vec{x})\, V(\vec{a},\vec{x})\, p(\vec{a}\mid \vec{x}).
  \label{eq:classical-obj}
\end{equation}
This expected payoff is linear in \(p\). We optimize the same expression over classical behaviors and a relaxation of the quantum set to obtain the optimum classical strategy and a quantum upper bound, respectively.

The expected payoff is linear over the convex polytope of classical behaviors, so an optimum is attained at a vertex corresponding to deterministic response functions. Shared randomness only mixes these responses, and the payoff of a mixture cannot exceed that of its best component.

Let $\mathcal{X}_i$ and $\mathcal{A}_i$ denote the sets of all possible local inputs and actions for the $i$th agent. Writing $f_i : \mathcal{X}_i \to \mathcal{A}_i$ for the $i$th agent's deterministic response, the classical optimum is
\begin{equation}
  C=\max_{f_1,\ldots,f_n}\;\sum_{\vec{x}} P(\vec{x})\, V\bigl((f_1(x_1),\ldots,f_n(x_n)),\vec{x}\bigr).
  \label{eq:classical-det}
\end{equation}

For $n$ agents with $m$ possible inputs and $k$ actions each, naive enumeration costs $O(m^n k^{mn})$ for a dense payoff tensor, treating each payoff lookup as constant time. Our implementation enumerates the responses of all but one agent and optimizes the remaining agent's action independently for each input, reducing this cost to $O(m^n k^{m(n-1)+1})$ while retaining the global optimum. The resulting deterministic responses provide the optimal classical strategy for comparison with quantum strategies. 


In a quantum strategy, the agents share an entangled state $|\psi\rangle$ in the joint Hilbert space $\bigotimes_{i=1}^n \mathcal{H}_i$. 
Given input $x_i$, the $i$th agent performs a local measurement on $\mathcal{H}_i$ with POVM elements $M_{a_i}^{x_i}$ producing an output $a_i$. According to the Born rule, the resulting  conditional probability distribution is:
\begin{equation}
\label{eq:qso-born}
p(\vec{a} \mid \vec{x}) = \text{Tr}\left[ \left(\bigotimes_{i=1}^nM_{a_i}^{x_i}  \right)|\psi\rangle\langle\psi| \right].
\end{equation}
The game specifies the input distribution and payoffs while the shared state and local measurements determine which behaviors the agents can realize. Each choice of shared state and local measurements gives a behavior $p$ and hence an expected payoff $\Pi[p]$. We denote the supremum of this payoff over all quantum strategies by $Q$, the \textit{quantum value} of the game.

Before searching for explicit quantum strategies, we compute an upper bound $Q_{\mathrm{ub}}$ on $Q$ using the NPA hierarchy~\cite{Navascues2008}.
At a given hierarchy level, the NPA semidefinite program maximizes the same payoff functional over a relaxation that contains all quantum behaviors, and therefore gives $Q_{\mathrm{ub}}\geq Q$. This semidefinite relaxation does not impose a bound on the Hilbert-space dimension. 
By dilating the local POVMs, we can represent them by projectors on enlarged local Hilbert spaces. 

Projectors belonging to different agents commute. 
Let $\mathcal{S}$ be a set of products of these projectors, including the identity. 
This set indexes a moment matrix whose entries correspond to the quantum expectation values of pairs of operators from $\mathcal{S}$. 
Every valid quantum realization produces a positive-semidefinite moment matrix. 
We impose this positivity, identity normalization, measurement completeness, projector orthogonality and idempotency, and commutation between the agents' operators. Maximizing the payoff function over the resulting feasible set gives an upper bound on the quantum value. Enlarging $\mathcal{S}$ gives tighter relaxations (see SI~\ref{app:sdp}).

A concrete example of the hierarchy levels is given below in the two-agent setting. At the first level, L1, the set $\mathcal{S}$ contains the identity and single measurement operators for both agents.
The intermediate level L1$+$AB adds cross-agent products such as $M_{a_1}^{x_1}M_{a_2}^{x_2}$ while L2 includes all possible operator words up to length two. Matrix entries can therefore involve products up to twice the maximum index-word length. 
The corresponding bounds satisfy
\begin{equation}
 Q_{\mathrm{ub}}^{\mathrm{L1}}
 \geq Q_{\mathrm{ub}}^{\mathrm{L1+AB}}
 \geq Q_{\mathrm{ub}}^{\mathrm{L2}}
 \geq Q_{\mathrm{ub}}^{\mathrm{L2+AB}}\geq Q.
 \label{eq:npa-ordering-results}
\end{equation}
The L2$+$AB  extension adds selected length-three cross-agent words such as $M_{a_1}^{x_1}M_{a'_1}^{x'_1}M_{a_2}^{x_2}$ and $M_{a_1}^{x_1}M_{a_2}^{x_2}M_{a'_2}^{x'_2}$. These terms are in addition to the cross-agent products already included at L2. 

We compare the exact classical optimum \(C\) with the calculated quantum upper bound \(Q^{L^{\ast}}_{\mathrm{ub}}\) and define the level-specific upper-bound gap as \(\Delta^{L^*}=Q^{L^{\ast}}_{\mathrm{ub}}-C\) where $L^{\ast}$ denotes the evaluated level of the NPA hierarchy. 
We choose a numerical tolerance $\epsilon>0$ according to the desired payoff precision, taking into account the solver's numerical accuracy. If $\Delta^{L^*} \leq \epsilon$, the calculated upper bound leaves no quantum-classical gap larger than the stated numerical tolerance. If \(\Delta^{L^{\ast}}>\epsilon\), the upper bound alone does not establish a resolvable advantage, but identifies a candidate regime in which we search for an explicit quantum strategy. The bound therefore rules out advantages above the numerical tolerance in unpromising regimes and focuses the search on games in which an advantage remains possible.

\subsection{Constructing quantum strategies}

We continue to the third step if a market-derived game has a positive upper-bound gap $\Delta^{L^*}$.
Here, we search for good quantum strategies that can realize the gap. 
We implement a see-saw strategy-construction procedure for market-derived nonlocal games, which we call the Quantum Strategy Search (QSS) algorithm.
QSS optimizes the shared state and local measurements directly from the input distribution and payoff tensor, without selecting a named Bell-game strategy in advance~\cite{WernerWolf2001,LiangDoherty2007,LiangLimDeng2009,PalVertesi2010}. 
As an alternative to this search method, batched gradient descent over the strategy parameterization has also recently been proposed \cite{xu2026efficientlyoptimizingquantumvalue}.
Our contribution is its integration with classical references, quantum upper bounds, numerical validation, and execution-level evaluation of the resulting strategies.  
Each feasible strategy gives a lower bound on the quantum value $Q$. An achieved payoff above the classical optimum $C$ establishes quantum advantage for the specified game. 
The QSS algorithm alternates between optimizing the shared state and each agent's local measurements, while holding the other components fixed. Since the expected payoff is linear in the shared state, a pure state suffices. For fixed local measurement operators, the payoff is $\bra{\psi}\Omega\ket{\psi}$ with 
\begin{equation}
\Omega=\sum_{\vec{x},\vec{a}}P(\vec{x})V(\vec{a},\vec{x})\bigotimes_{i=1}^{n}M_{a_i}^{x_i}.
\label{eq:qss-payoff-operator}
\end{equation}
The operator $\Omega$ is Hermitian, so the optimal state update is its normalized eigenvector with the largest eigenvalue~\cite{LiangLimDeng2009}. Replacing the current state with this eigenvector cannot decrease the payoff. 
Next, we fix the shared state and all agents' measurements except those of the $i$th agent. The expected payoff then becomes linear in the $i$th agent's POVM elements. 
Specifically, the expected payoff can be written as:
\begin{equation}
\sum_{x_i, a_i} \text{Tr}\left(M_{a_i}^{x_i} G_{a_i}^{x_i}\right),
\end{equation}
where the effective environment operator $G_{a_i}^{x_i}$ absorbs the state, the other agents' measurements, the input distribution, and the game payoffs:
\begin{equation}
G_{a_i}^{x_i} = \sum_{\vec{x}_{-i}, \vec{a}_{-i}} P(\vec{x})V(\vec{a},\vec{x}) \operatorname{Tr}_{-i} \left[ |\psi\rangle\langle\psi| \bigotimes_{j \neq i} M_{a_j}^{x_j} \right].
\label{eq:qso-local-score-updated}
\end{equation}
Here, $\operatorname{Tr}_{-i}$ denotes the partial trace over all systems except that of agent $i$. Because the POVM elements $M_{a_i}^{x_i}$ must be positive semidefinite and sum to identity for each input $x_i$, finding the optimal measurement for agent $i$ reduces to solving an independent semidefinite program (SDP). 
We update each agent's measurements in turn. For binary actions, the measurement update has an analytical solution; for more than two actions, we solve an SDP.
Each exact update does not decrease the payoff, so the sequence of payoff values converges. This does not guarantee convergence of the state and measurements to a single point or attainment of the global maximum. We therefore use multiple random initializations and retain the candidate with the highest payoff. Increasing the local dimension embeds the strategies available at lower dimensions, but need not improve the result of a finite QSS. 

We compare the achieved payoff with the NPA upper bound to assess how closely the strategy approaches the quantum value. To ensure the search never returns a payoff below \(C\), we retain the optimal classical strategy as a candidate ansatz. Once a strategy is selected, its implementation requires preparing the shared entangled state and having each agent apply the local measurement specified by its input.

\subsection{End-to-end evaluation}
\label{sec:end-to-end}
The final stage is to backtest the quantum strategy identified in stage 3. The preceding stages operate on the abstracted tensors \((P,V)\). 
A quantum advantage in the extracted game does not by itself establish an advantage in market execution, where latency, competition for available liquidity, partial fills, and other trading frictions can affect realized payoffs. We therefore evaluate the strategies in the same backtesting engine used to build the game, testing whether the predicted gains persist under some of these execution effects and finite-sample fluctuations, within the limitations of historical replay.
\begin{figure}[t]
    \centering
    \includegraphics[width=\columnwidth]{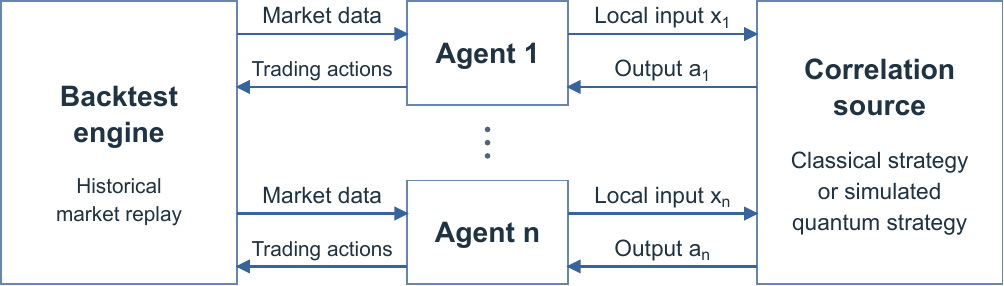}
    \caption{Architecture of the end-to-end backtest. 
    The engine exchanges
    quotes and orders with each agent as in the uncoordinated baseline; to
    decide whether to act on an opportunity, each agent queries a shared
    correlation source with a request and receives a signal. 
    The source supplies the coordination resource, e.g., classical randomness, or measurements on a shared entangled state.}
    \label{fig:coined-backtest}
\end{figure}

Figure~\ref{fig:coined-backtest} depicts the multi-agent backtest architecture using the \texttt{NautilusTrader} backtesting engine. The engine streams quotes to each agent and executes the orders they return, as in the uncoordinated baseline. Our extension changes how an agent decides whether to act on a detected opportunity. In the simulation, each agent’s local observation is used to generate an output from the prescribed joint behavior, and each agent receives only its own output. The simulation does not pass one agent’s input or outcome to another. For \(n\) agents, the same procedure uses a shared \(n\)-partite state and returns one local output to each agent. In a physical implementation, each agent would obtain its output by locally measuring its part of the distributed state, using the same input-to-measurement and outcome-to-action rules.
The detailed implementation is described in SI~\ref{app:coordinated-backtest}.

In the backtest simulation, each window of Sec.~\ref{sec:nonlocal-game} is assigned its own shared state, so every round of the game is played with a fresh state.
An agent identifies the window of a trading signal from the publishing venue's timestamp on the quote that generates it, so feed latency and the agents' local clocks do not affect which window an observation belongs to, and the agents need no communication to agree on it.
Each agent measures its part of the state for the first trading signal it observes in the window and acts according to the resulting output.
If it observes no signal, it takes no action.
Since only one shared state is assigned per window, each agent forgoes subsequent opportunities on other assets in that window.
Agents running classical strategies are not held to this rule in the backtest and act on every signal for their assets.
This is a departure from the one-input restriction under which the \((P,V)\) tensors are computed, and so it is part of what the validation step tests.
The comparison between quantum and classical strategies must also account for rounds in which only some agents observe a trading signal, without post-selecting successfully matched events.

Fixed boundaries come at a cost since two related signals, such as a single price movement observed at two venues, separated by a delay $\Delta t$ can fall in different windows, with probability of order $\Delta t / w$.
In this case the agents draw on different shared states, and their decisions on that pair of signals are uncorrelated.
A wider window makes such splits rarer, so coordination acts more reliably on related signals.
Against this, each agent using quantum coordination takes a single input per window, so with a wider window it forgoes more of the signals on other assets that follow its first.

The width also determines what the game represents.
A window groups the signals that the game treats as simultaneous, so it should match the timescales over which the market effects captured by the game, such as the lifetime of a price dislocation, actually act.
A wider window also makes it more likely that an agent observes signals on more than one asset within one window, which a single input cannot represent.

In the example below, we use windows of $w = 100$~ms. Among windows in which an agent observes a signal, fewer than $0.3\%$ contain signals on multiple assets on every modified record (SI~\ref{app:pv-extraction}).
This window sets the timescale for distributing and matching entangled states. It does not give agents 100~ms to delay their actions, since each must commit immediately after observing an opportunity to secure queue priority.

We run the constructed quantum strategy and the classical comparator on the same market data in the same engine, and compare their realized payoffs both with each other and with the game-theoretic values predicted from \(P,V\).
The first comparison gives the quantum--classical difference under the backtest's modeling assumptions; the second tests how much of the separation established on the tensors survives execution.

\section{Numerical experiments and results}
\label{sec:realistic-games}

We use a two-agent, three-stock cross-venue arbitrage setting to benchmark the market-derived-game approach. Each agent monitors a pair of venues for opportunities in the three stocks, trading when the best bid on one venue exceeds the best ask on the other. If the agent acts, it submits paired orders to buy at the lower ask and sell at the higher bid. The setting shares the broad structure of Brandenburger and La Mura~\cite{blm2015}, but it gives rise to a different game. We define the agents’ signals and available actions from a financially grounded trading strategy, derive the input probabilities from historical order-book data, and obtain the payoffs by backtesting each joint action. The resulting signals, action spaces, probabilities, and payoffs are tied to observed market behavior. This also lets us identify and quantify concrete interactions, such as competition for shared liquidity and cross-stock repricing, that shape the potential for quantum coordination.

An agent's two orders may fill in unequal quantities, leaving a residual long or short position. To limit exposure to subsequent price movements, the strategy closes any residual position through an offsetting trade, a process referred to as \emph{unwinding}. Coordination determines whether the agent initiates the arbitrage trade, while the underlying execution and unwinding rules remain the same. We consider decisions that must be made before the agents can exchange information about their current opportunities~\cite{blm2015,Wissner2010}.

We divide the historical timeline into non-overlapping windows of width $w$, with boundaries agreed upon in advance. Each nonempty window defines one round of the game and windows in which neither agent observes a trading signal are discarded. For each agent, $x_i\in\{0,1,2,3\}$ identifies the stock opportunity observed in a window with $x_i=0$ denoting no opportunity. The action $a_i=1$ means trade and $a_i=0$ means yield. The coordinated strategy takes one input per agent per window and acts on the first qualifying signal as described in Sec.~\ref{sec:framework-overview}. 

The observed inputs determine the distribution $P(\vec{x})$ normalized over nonempty windows. To obtain the payoff tensor $V$, we replay the historical record under each allowed joint action and average the realized payoff over windows with the corresponding input tuple. Thus, $V$ includes stock-dependent payoffs, partial fills, and unwinding.  

We use \texttt{Databento} Level-1 quotes and trades~\cite{databento}. Stock selection was guided by the observation that highly liquid stocks tended to offer smaller price discrepancies, whereas infrequently traded stocks produced too few opportunities for simultaneous agent activity.
We chose the three stocks: AMD, INTC, and MU, traded on three venues: NYSE Arca (\code{ARCX}, Mahwah), Cboe BZX (\code{BATS}, Secaucus), and Nasdaq (\code{XNAS}, Carteret).
Agent~A monitors \code{ARCX}--\code{BATS} and agent~B monitors \code{ARCX}--\code{XNAS} for all three stocks.
The agents compete for liquidity at \code{ARCX}. We chose high-VIX days because larger price fluctuations were expected to yield more opportunities requiring coordination; in particular, we use the twenty highest-VIX days in our data, spanning four volatility episodes from August 2024 to March 2026. These choices reflect exploratory observations rather than a formal optimization of stocks or dates. The venues are tens of kilometers apart, giving inter-venue communication delays on the order of a hundred microseconds or more~\footnote{\href{https://www.zayo.com/resources/route-overview-new-jersey-triangle/}{Zayo, Route Overview: New Jersey Triangle}.}. Such delays can make waiting for coordination costly when trading opportunities are short-lived.

In these experiments, an opportunity must offer a price discrepancy of at least one tick, and both orders are sized to the smaller displayed quantity at the relevant bid and ask. Realized profit includes partial fills and any unwinding costs, with fills affected by available liquidity and feed latency where it is nonzero. Exchange fees are set to zero. As the uncoordinated reference, a greedy agent acts on every opportunity it detects, without price prediction or deliberate inventory holding. 

\begin{figure*}[t]
    \centering
    \includegraphics[width=\textwidth]{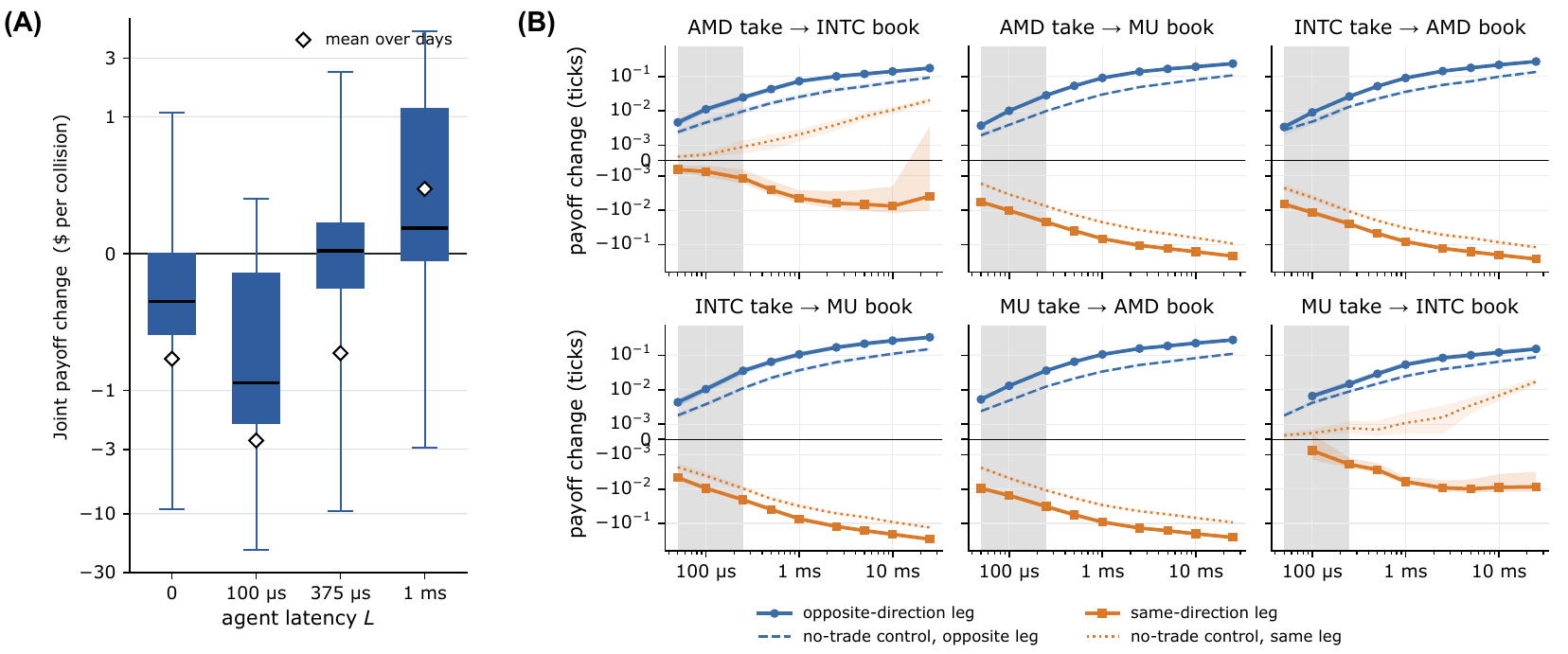}
    \caption{Collision costs and potential cross-stock synergy for AMD, INTC, and MU. (A) Change in combined payoff per same-stock collision when liquidity tracking is enabled versus disabled, as a function of feed latency. Negative values indicate losses from liquidity competition. Boxes show the interquartile range across twenty trading days; black lines, whiskers, and white diamonds indicate the median, range, and mean. (B) Change in the quoted payoff of a subsequent trade in a second stock following an aggressive trade in the first, relative to pre-trade quotes and measured in ticks of the second stock. Blue and orange curves represent opposite- and same-direction trades. Solid curves follow aggressive trades; dashed blue and dotted orange curves follow quote changes without trades. The blue trade-minus-control difference gives an optimistic estimate of potential synergy, assuming full attribution and execution; adverse same-direction repricing adds a penalty. Shading shows $68\%$ intervals from a joint bootstrap over trading days; control bands are generally narrower than the lines.}
    \label{fig:synergy-and-penalty}
\end{figure*}

Our backtests use \texttt{NautilusTrader} v1.224.0, on data retrieved with the \texttt{Databento} Python client v0.74.0 from the \code{XNAS.ITCH}, \code{ARCX.PILLAR}, and \code{BATS.PITCH} datasets (\code{mbp-1} and \code{trades} schemas), covering the regular trading hours, 09:30--16:00 New York time, on each date.
Experiments use window widths $w\in\{1, 10, 100\}$~ms and venue-to-agent feed latencies $L\in\{0, 50, 100, 375, 1000\}\,\mu\mathrm{s}$. The following results specify the settings used for each analysis.

With the trading rule and local signal fixed, the coordination resource can help by changing the distribution of the agents' joint actions following simultaneous trading signals. Whether this improves performance depends on how the combined payoff varies across those joint actions. The measurements below examine how competition for shared liquidity and cross-stock repricing affect the payoff.

Liquidity competition occurs when one agent's order consumes the liquidity needed by another agent at the same venue. We estimate the resulting collision cost by comparing two-agent greedy backtests with liquidity depletion enabled and disabled. Disabling depletion provides a non-interacting reference while enabling it introduces competition for the remaining displayed volume. Unequal fills leave residual positions that are unwound at the best available quote. We count same-stock order submissions within $w_{\mathrm{col}}=1\,\mathrm{ms}$ as collisions and measure the change in combined profit per collision with a negative change corresponding to a collision cost. Here $w_{\mathrm{col}}$ is only the interval for identifying collisions in this measurement and is distinct from the game-window width $w$. To examine how delayed market information affects competition, we scan feed latencies from zero to $1\,\mathrm{ms}$. At low latency, shared-book updates can provide indirect feedback about consumed liquidity, while at longer latencies, stale quotes increasingly expose agents to liquidity already taken by other participants. The largest measured cost occurs at $L=100\,\mu\mathrm{s}$ with a cost of $\$2.6$ per collision corresponding to about $60\%$ of the non-interacting payoff (Fig.~\ref{fig:synergy-and-penalty}(A)). At $L=1\,\mathrm{ms}$, depletion tracking instead improves the mean payoff, consistent with preventing attempted trades against already-consumed liquidity. The corresponding difference remains small in the single-agent control.

Cross-stock repricing occurs when price changes in one stock are associated with changes in another, altering the payoff available to a subsequent trade. When the stocks move together, selling one after buying the other can benefit from the price increase, whereas a subsequent purchase becomes more expensive. To estimate this effect, we examine isolated aggressive trades in the first stock at \code{ARCX} and measure the change in the quoted payoff of a subsequent leg in the second stock after a horizon $\tau$. Quote changes in the first stock without a trade provide a control for repricing without the initiating trade. For responding books with spreads of at least two ticks, the mean favorable change at $\tau=100\,\mu\mathrm{s}$ is approximately $0.01$ ticks, roughly twice the control (Fig.~\ref{fig:synergy-and-penalty}(B)). We treat the excess over this control as an optimistic estimate of potential synergy, since this assumes it is caused by the initiating trade and can be captured in execution. Common information and differences between the event samples may inflate or mask the causal effect, while execution frictions can reduce the realized benefit.

These measurements provide reference scales for the interaction terms in $V$, while $P$ determines how frequently the corresponding input pairs occur. 
Liquidity competition and cross-stock repricing involve dynamic market responses that a fixed historical record cannot fully capture. These estimates therefore provide approximate scales for the interactions, rather than precise predictions of their effects in live trading. The market-derived payoff tensors constructed from the three-stock data analyzed here show no resolved quantum-classical gap. In the next section, we study a simplified version of these games by varying the joint payoff effects associated with collisions and cross-stock trades to estimate the scale needed for a quantum advantage.

\subsection{Controlled joint payoffs and quantum advantage}
\label{sec:npa-results}

To determine what joint payoff changes could produce an advantage, we consider a simplified game on the input distribution $P$ extracted from a modified historical market data tape from April 9, 2025. We use a common single-agent payoff equal to the average payout over the six solo-input cells of the original tape. When both agents trade, we add a fixed joint-payoff increment for trading different stocks and apply a collision penalty for trading the same stock. The scan varies these two joint-payoff adjustments while holding $P$ fixed. 

The modified tape is constructed by adding synthetic quote events to increase the frequency of joint opportunities, particularly different stock opportunities. For April 9, the modified tape has 2,221 nonempty rounds out of which 359 were from the original record and 1,862 were injected. The resulting $P$ has approximately 20\% solo events (only one agent sees a trading signal), 20\% same-stock joint events, and 60\% different-stock joint events, with the joint opportunities symmetrized across stock pairs. The full injection and extraction procedure is described in SI~\ref{app:backtest-mods}.

\begin{figure}[t]
    \centering
    \includegraphics[width=\columnwidth]{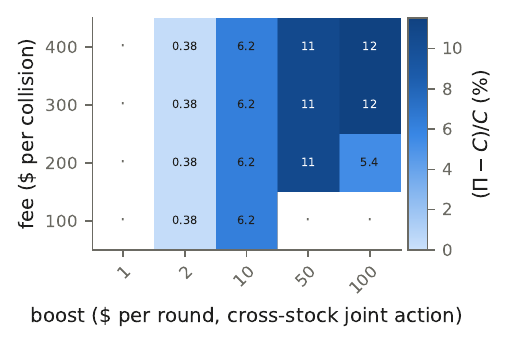}
    \caption{Relative quantum advantage $(Q_{\mathrm{ach}}-C)/C$ as a function of the boost, paid per round in which the two agents act on different stocks, and the fee, charged per collision (both agents acting on the same stock), where $C$ is the classical optimum and $Q_{\mathrm{ach}}$ the best two-qubit QSS payoff (denoted by $\Pi$ in the plot).
    All cells use the same modified April 9, 2025 order-book data, sharing its input distribution and trading payouts, differing only in the boost and fee.
    The cell at boost \$50, fee \$200 is the game played in Fig.~\ref{fig:quantum-backtest}.
    Dots mark games with no achieved advantage to numerical precision.
    The modified record is described in SI~\ref{app:backtest-mods}.}
    \label{fig:doped-gap-scan}
\end{figure}

For each replayed joint action, we adjust the combined payoff by one penalty when both agents trade the same stock, or by one joint-payoff increment when they trade different stocks. These post-simulation adjustments provide a controlled model of the market interactions. Figure~\ref{fig:doped-gap-scan} shows the achieved relative advantage $(Q_{\mathrm{ach}}-C)/C$ across the scan. Here, $Q_{\mathrm{ach}}$ is the best two-qubit QSS payoff found for each game. A point with $Q_{\mathrm{ach}}>C$ therefore has an explicit quantum strategy with the NPA bound giving an upper bound on how large that advantage could be. The scan shows that the joint-payoff increment and collision penalty must be sufficiently large for quantum coordination to help. When the penalty is small relative to the increment, a classical strategy that accepts collisions can be favorable. Increasing the penalty makes disjoint stock assignments preferable and opens a regime in which quantum coordination performs better.

At the interaction scales measured in Fig.~\ref{fig:synergy-and-penalty}, no gap opens, even on the modified record: the collision penalty of $\$2.6$, measured at $100\,\mu\mathrm{s}$, and the $\$0.01$ joint-payoff increment, an indicative gross repricing scale, both lie below the range of Fig.~\ref{fig:doped-gap-scan}, whose smallest boost, $\$1$, yields no advantage at any fee.
The QSS strategies used in the scan are explicit two-qubit strategies.
Comparisons of finite NPA levels and details of the QSS implementation are given in SI~\ref{sec:npa-convergence} and SI~\ref{app:see-saw}, respectively.

\subsection{End-to-end backtesting results}
\label{sec:q_backtest_results}
\label{sec:backtest-results}

The scan identifies payoff regimes in which the quantum strategy outperforms the best classical strategy.
 Here, we perform backtesting to examine whether the advantage survives when agents make decisions on a sequence of quotes and submit orders as in realistic trading environments. We run the quantum strategy and the best classical strategy on the same modified tape and reconstruct the payoffs from their trade reports. We repeat the quantum backtesting with different sampled measurement outcomes to estimate the resulting variability. 

\begin{figure*}[t]
    \centering
    \begin{subfigure}{0.45\linewidth}
    \includegraphics[width=\linewidth]{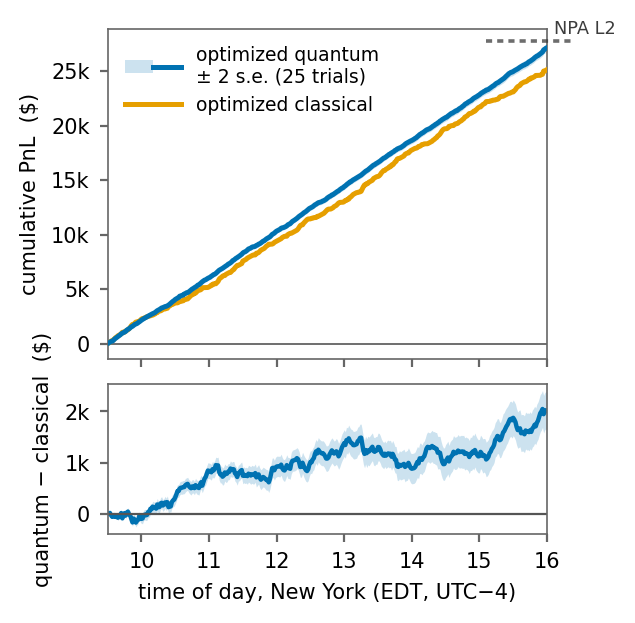}
    \caption{}
    \label{fig:quantum-backtest}
    \end{subfigure}
    \begin{subfigure}{0.52\linewidth}
    \includegraphics[width=\linewidth]{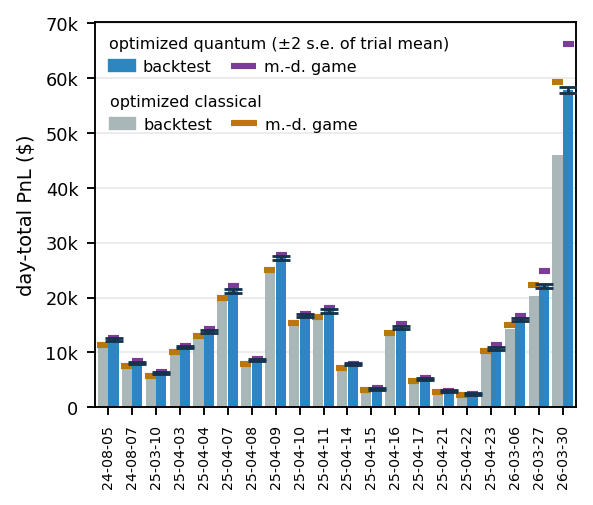}
    \caption{}
    \label{fig:quantum-backtest-20day}
    \end{subfigure}
    \caption{(a) Backtest on the modified record of April 9, 2025, with joint-payoff increment $\$50$ and collision penalty $\$200$ (see SI~\ref{app:backtest-mods}).
    Upper panel: cumulative PnL of the optimized quantum (mean of 25 trials, shading $\pm 2$ s.e.) and classical strategies; the dashed line is the NPA level-2 bound of the market-derived game, which the optimized quantum strategy attains in the game to within $0.001\%$.
    Lower panel: quantum minus classical PnL.
    (b) Daily PnL over twenty modified records: results of backtests (bars), and payoffs in the market-derived (m.-d.) game computed from $(P,V)$ tensors (ticks).
    The $(P,V)$ payoff of the optimized quantum strategy lies within $0.0022\%$ of the NPA level-2 bound on every day; the backtests track the game values, falling short on the most active days (see SI~\ref{app:backtest-reconciliation}).}
    \label{fig:end-to-end-backtest}
\end{figure*}

We use the controlled tape and joint payoff adjustments described in the previous section. Figure~\ref{fig:quantum-backtest} shows the backtesting results for the April 9, 2025 modified session. The mean final quantum payoff exceeds the classical payoff by $8\%$. The uncertainty reflects variation across repeated backtests that differ only in the sampled quantum measurement outcomes. This shows that the game-level advantage can survive execution on the modified tape. 

Figure~\ref{fig:quantum-backtest-20day} compares game-level predictions with backtested payoffs across twenty modified daily tapes. The replayed payoffs generally track the predictions and retain the quantum strategy's higher payoff. On the most active days, replayed payoffs fall below their game-level predictions because the backtest's stale-liquidity guard suppresses orders against historical quotes whose displayed liquidity has already been consumed by an agent. 
An injected quote can create an apparent opportunity against a historical quote whose liquidity the agent has already consumed. Suppression is strongest when many injected opportunities use the same historical quote before it is refreshed.
This discrepancy arises from the interaction between the synthetic opportunity injection and replaying an unchanged historical quote stream. It should not be interpreted as evidence that the same suppression occurs at this rate in unmodified market data (see SI~\ref{app:backtest-reconciliation}).

Together, these results show that a quantum strategy identified from the market-derived ($P,V$) tensors can retain an advantage over the best classical strategy when its decisions are turned into orders and run through backtesting.  

\section{Conclusion and Outlook}
\label{sec:conclusion}

We have introduced market-derived nonlocal games, a new systematic approach in which financially grounded trading strategies and historical market data determine the game to be tested. The base trading strategy specifies each agent's signals and available actions based on market dynamics and financial objectives. Historical data and backtests then determine the agents’ signals and the payoffs associated with their actions. This lets us compare the best classical strategy with quantum upper bounds, search for entangled strategies, and test their payoffs in end-to-end backtests.

In the cross-venue arbitrage study, we use historical order-book data to estimate the payoff effects of liquidity collisions and cross-stock repricing. At the measured scales, we find no resolved quantum-classical gap. We then test controlled benchmarks with stronger interactions and more joint opportunities. In those settings, our search finds quantum strategies that outperform the best classical strategy, and end-to-end backtests show that their advantage persists. This example shows how market-derived games can connect observed market behavior to the search for quantum coordination strategies.

The cost of finding classical optima and quantum upper bounds increases with the number of agents, signals, actions, and NPA hierarchy level. But the size of a game also depends on how the trading decision is represented. Restricting the local actions to choices relevant to the strategy can keep a game manageable while retaining the market interactions of interest. Additional structure in the games can also be exploited to lower the cost. For example, dynamic programming has been used to compute classical bounds for structured multipartite Bell inequalities \cite{Tura2017}, while symmetry reduction can reduce the size of NPA relaxations \cite{Ioannou2022}. Developing similar methods for market-derived games could make it possible to analyze larger strategy and market combinations.

The two-agent, three-stock cross-venue arbitrage example is only one possible market-derived nonlocal game. Each financially grounded trading strategy and market environment defines a different game. Prediction-based strategies~\cite{kearns2013machine} and strategies designed around risk exposure or variance~\cite{cartea2015risk,fodra2012}, for example, could be used to ask whether quantum coordination improves decisions that are already determined by financial and market considerations. This provides a systematic way to test quantum coordination across trading strategies, stocks, and market conditions using games derived from observed market behavior.

Once a strategy shows an advantage in the market-derived game and retains it in backtesting, the next step would be to test the coordination on a live signal feed. Entanglement distribution between quantum memories over deployed fiber has already been demonstrated at distances relevant to the venues considered here, including silicon vacancy-based memory nodes connected over 35 km~\cite{knaut2024} and nitrogen-vacancy memory nodes over 25 km~\cite{stolk2024}. While these demonstrations have not yet met the distribution rates necessary, it is important to note that the local measurement time and the entanglement distribution rate need not be the same timescale. In some candidate systems, local measurements take tens of microseconds, which are comparable to the decision times considered here~\cite{knaut2024}. Entangled pairs can instead be distributed in advance and held in quantum memories until a coordination signal arrives. 
If one entangled pair is consumed per nonempty coordination window, even a window as short as 1 ms requires at most 1,000 pairs per second; longer windows or fewer occupied windows reduce this requirement.

\section*{Use of AI Tools}
AI tools assisted with manuscript wording and phrasing, code development for numerical simulations, HPC execution and plotting, and preparation of supplementary tables. Tools used included Claude Fable 5.1, Sonnet 5, and Opus 5.5; ChatGPT GPT-6 Astra; and Google Gemini 3 Flash, 3.5 Flash, and 3.6 Flash. The authors developed the scientific framework, specified the computational tasks, and interpreted the results. They carefully checked the reported numerical results, underlying intermediate outputs, figures, and tables for correctness and consistency.

\begin{acknowledgments}
We thank Marco Pistoia, Chad Sakac, Dawei Ding, Liang Jiang, John Gardiner, and Isaac H. Kim for valuable discussions and insights related to this work.     
\end{acknowledgments}

\nocite{*}
\bibliography{references}

\clearpage
\onecolumngrid
\begin{center}
{\large\bfseries Supplementary Information}\par\medskip
{\bfseries Testing Quantum Advantage in High-Frequency Trading with Market-Derived Nonlocal Games}
\end{center}
\medskip
\twocolumngrid
\renewcommand{\theHsection}{SI.\arabic{section}}
\renewcommand{\theHsubsection}{\theHsection.\arabic{subsection}}
\renewcommand{\theHequation}{SI.\arabic{equation}}
\renewcommand{\theHfigure}{SI.\arabic{figure}}
\renewcommand{\theHtable}{SI.\arabic{table}}

\setcounter{section}{0}
\setcounter{equation}{0}
\renewcommand{\theequation}{S\arabic{equation}}
\setcounter{figure}{0}
\renewcommand{\thefigure}{S\arabic{figure}}
\setcounter{table}{0}
\renewcommand{\thetable}{S\arabic{table}}

\section{Constructing market-derived nonlocal games}
\label{app:p-extraction}
\label{app:pv-extraction}
\label{app:backtest-mods}
Here, we follow the historical market data from the engine's output to the games used in the backtests, illustrated on April 9, 2025, the day of Fig.~\ref{fig:quantum-backtest}.
The same extraction is applied twice: first to the original record, where it tests the representation and supplies the starting composition of opportunities, and then to the modified record, whose output is the realized tensor used throughout.

In the record, an arbitrage opportunity has a lifetime.
It opens with the quote tick at which the highest bid at one venue first exceeds the lowest ask at the other, $p_{\text{bid}} > p_{\text{ask}}$, persists while subsequent ticks leave the two prices crossed, and closes with the tick that uncrosses them.
We call this span a \emph{crossing interval}.
The engine reports a detection at every quote tick within it, so the same opportunity is typically detected several times.
These detections are merged into a single event per crossing interval; events are grouped into the coordination windows that define the rounds of the game; and rounds are binned into the cells $(x_1,x_2)$ of $P$.
Only nonempty windows, $(x_1,x_2)\neq(0,0)$, enter the normalization of $P$ and the round counts reported below.
The payoffs $V$ are measured on the same rounds (Sec.~\ref{sec:nonlocal-game}).
The engine replays each record with only agent~1 acting, only agent~2 acting, and both acting.
The extracted value for $V(a_1,a_2,x_1,x_2)$ is then the mean realized profit under the actions $(a_1,a_2)$, taken over all rounds with input $(x_1,x_2)$.

The same accounting also tests how well the game's input alphabet captures the market.
Each round compresses the market signals an agent sees within one window into a single input: one stock symbol, or none.
We use two checks of the construction itself.
The input distribution $P_{\text{first}}$, estimated from the first detection of each event, agrees with $P_{\text{all}}$, estimated from every detection, to within a total-variation distance $\tfrac12\sum_{x_1,x_2} |P_{\text{first}}(x_1,x_2) - P_{\text{all}}(x_1,x_2)|$ of at most $0.004$ on every original record, so the resulting $P$ does not depend on which detection is kept.
Also, the extracted $P$'s entries sum to one, so every round falls in exactly one cell.

A third check measures what the compression discards, namely, the windows in which an agent sees opportunities on more than one symbol, which a single input cannot represent.
We define $\xi$ as the probability that an agent sees opportunities on two or more symbols in a window, conditioned on its seeing at least one.
Such windows are rare: $\xi = 0.0065$ pooled over the original records, and the injected rounds add none.
When the six classical partitions are scored on the same order book with and without the single-input restriction, the best unrestricted partition collects at most $0.17\%$ more than the best restricted one on any modified record, and no more at all on nine of the twenty (Table~\ref{tab:extraction}).
The restricted value is the like-for-like benchmark, since both strategy classes then play the same market-derived game; the unrestricted value measures how well that game fits the record at this window width.
On the original books, where such windows are a larger share of fewer rounds, the excess exceeds $1\%$ on six days and reaches $22\%$ on the smallest.
A large $\xi$ is not an obstacle to the approach but a sign that the game should be enlarged, for example to quantum strategies in which each agent holds more than one qubit, so that it can act on more than one symbol per round.

The extraction of the original record also gives its composition of opportunities, the ``original rounds" columns of Table~\ref{tab:recipes}.
In this constructed game, coordination matters only in windows in which both agents observe an opportunity, and the quantum advantage arises only when those opportunities are on different assets; the historical records contain almost none of these.
For the controlled end-to-end tests, we therefore inject synthetic quote events into the historical record to increase their frequency.
Our pilot procedure places events uniformly throughout the trading day, though more sophisticated distributions are realizable.

The construction imposes the symmetries of the Brandenburger and La Mura game~\cite{blm2015} on the joint opportunities, while the relative frequency of each class of opportunities is chosen to make the quantum--classical gap clearly visible.
Within each class, every pair of symbols is equally represented, so neither agent and no symbol is given a preferred role.
We increase the number of different-stock joint opportunities, which qualify for a synergy boost; since the quantum advantage arises only from these opportunities, their frequency, together with the size of the boost, sets the magnitude of the gap~\cite{blm2015}.
At a finite collision fee, same-stock joint opportunities must also be sufficiently frequent to keep collision-prone strategies from becoming favorable.

Single-agent opportunities make up the rest of the modified record and retain the original balance between agents. The imposed symmetry between agents therefore applies only to joint opportunities.

The ratio of injected to original events is therefore set by this target composition and by how much of each class the original record already supplies, rather than being fixed independently.
The twenty days of Fig.~\ref{fig:quantum-backtest-20day} share the same fraction of opportunities in each class, but their original records differ in how many opportunities they contain, so the number of injected events, and its ratio to the original ones, differs from day to day.
Table~\ref{tab:recipes} lists the original and injected rounds of each class for every day in Fig.~\ref{fig:quantum-backtest-20day}.
The original columns are the class counts of $P$ extracted from each original record, and the original and injected columns together are those of the modified record.

\begin{table}[t]
\caption{Record-modification recipes for the twenty days of Fig.~\ref{fig:quantum-backtest-20day}.
$\hat m$ is the mean single-agent payout per round on the original record, averaged with equal weight over the six cells $(x,0)$ and $(0,x)$ of $P$, and is used to size the injected events.
The \emph{original} columns count the rounds of each class already present in the historical order book, and the \emph{injected} columns the rounds added by the recipe.
On every day the modified order book has class fractions $0.2$ (solo), $0.2$ (agents see same-stock) and $0.6$ (agents see different-stock), to within $0.01$ from the rounding of the counts.
The backtests for all days were performed at zero feed latency with a joint-payoff increment of $\$50$ and a collision penalty of $\$200$.
}
\label{tab:recipes}
\begin{ruledtabular}
\begin{tabular}{lrrrrrrr}
 & & \multicolumn{3}{c}{original rounds} & \multicolumn{3}{c}{injected rounds} \\
\cline{3-5} \cline{6-8}
date & $\hat m$ (\$) & solo & same & diff & solo & same & diff \\
\hline
2024-08-05 & 0.70 & 149 & 27 & 0 & 65 & 186 & 642 \\
2024-08-07 & 0.25 & 94 & 8 & 0 & 54 & 139 & 438 \\
2025-03-10 & 0.35 & 61 & 3 & 0 & 49 & 108 & 330 \\
2025-04-03 & 1.41 & 98 & 14 & 0 & 81 & 163 & 534 \\
2025-04-04 & 0.96 & 164 & 18 & 1 & 73 & 219 & 707 \\
2025-04-07 & 1.03 & 295 & 17 & 1 & 69 & 346 & 1091 \\
2025-04-08 & 1.20 & 132 & 14 & 1 & 10 & 127 & 425 \\
2025-04-09 & 1.36 & 315 & 43 & 1 & 130 & 401 & 1331 \\
2025-04-10 & 1.54 & 245 & 24 & 0 & 28 & 249 & 816 \\
2025-04-11 & 1.92 & 274 & 18 & 0 & 12 & 267 & 852 \\
2025-04-14 & 0.52 & 107 & 9 & 0 & 28 & 126 & 408 \\
2025-04-15 & 0.70 & 52 & 4 & 0 & 7 & 56 & 180 \\
2025-04-16 & 0.44 & 107 & 20 & 0 & 154 & 241 & 780 \\
2025-04-17 & 2.44 & 74 & 9 & 0 & 7 & 72 & 240 \\
2025-04-21 & 0.35 & 44 & 4 & 0 & 8 & 47 & 156 \\
2025-04-22 & 0.71 & 29 & 7 & 0 & 12 & 35 & 126 \\
2025-04-23 & 0.57 & 94 & 15 & 0 & 98 & 177 & 576 \\
2026-03-06 & 0.40 & 266 & 24 & 1 & 21 & 264 & 863 \\
2026-03-27 & 0.57 & 353 & 32 & 0 & 73 & 394 & 1278 \\
2026-03-30 & 0.43 & 812 & 92 & 1 & 333 & 1051 & 3431 \\
\end{tabular}
\end{ruledtabular}
\end{table}

The modified record is passed through the same extraction, and its $P$ reproduces the target class fractions to within $0.01$ on all twenty days, consistent with rounding the event counts (Table~\ref{tab:recipes}).
Table~\ref{tab:extraction} compares the two extractions.
On April 9 the modification raises the number of rounds from 359 to 2,221 and occupies all 15 cells of $P$, against 9 on the original record.
It also changes the character of the reduction.
An injected opportunity is a single quote and is detected about once, against about five times for a crossing on the original record, so on the modified records an event is detected $1.2$ to $1.8$ times, and a round contains $1.82$ to $1.89$ events on every day, since most injected rounds are observed by both agents.
On every modified record $\xi$ is at most $0.0027$, below the level of $0.006$, because each injected round carries a single symbol per agent by construction and dilutes the original rounds rather than altering them.

\begin{table*}[t]
\caption{Coarse-graining of the engine output into the input distribution $P$ for 2025-04-09, the day of Fig.~\ref{fig:quantum-backtest}, with the range over all twenty days (modified order books unless noted).
Detections of an opportunity are merged into one event per crossing interval, events are grouped into coordination rounds, and rounds are binned into the cells $(x, y)$ of $P$.
The checks compare $P$ estimated from the first detection of each event with $P$ from every detection, verify the normalization of $P$, and give the probability $\xi$ that an agent sees two or more symbols in a nonempty round.
The last two rows score the six classical partitions on the same record with and without the single-input restriction and give how much the best unrestricted partition exceeds the best restricted one, as a percentage of the latter.}
\label{tab:extraction}
\begin{ruledtabular}
\begin{tabular}{llrr}
stage & quantity & 2025-04-09 & range, 20 days \\
\hline
engine output & detections (both agents) & 6699 & 578--15240 \\
events & distinct crossing intervals & 4194 & 389--10709 \\
 & detections per event & 1.60 & 1.21--1.79 \\
rounds & coordination windows & 2221 & 209--5720 \\
 & events per round & 1.89 & 1.82--1.89 \\
cells & occupied cells of $P$ (of 15) & 15 & 13--15 \\
\hline
checks & total-variation distance, first vs.\ all detections & 0 & 0--0.0009 \\
 & normalization error of $P$ & 0 & $<10^{-15}$ \\
 & $\xi$, modified order book & 0.0008 & 0--0.0027 \\
 & $\xi$, original order book & 0.0074 & 0--0.032 \\
 & classical optimum gained without input restriction, modified order book & 0.09\% & 0--0.17\% \\
 & \hspace{19em} same, original order book & 4.1\% & 0--22\% \\
\end{tabular}
\end{ruledtabular}
\end{table*}

After each simulation, we extract the trade reports and use these to compute a modified PnL: each coordination window in which both agents trade receives a single collision penalty if they traded the same stock, or a single synergy boost if they traded different stocks.
This supplies a controlled model of market interactions that replay of a fixed historical record cannot capture.

The modifications to the PnL can be interpreted as fees for contending at the shared venue and rebates for coordinating across assets, but they do not reconstruct the market's response to the simulated orders.
It is in effect a mean-field approximation to the underlying market dynamics, replacing that response by a constant for each class of joint opportunity.
As the market's response enters only through these two constants, the conditions for a quantum advantage become bounds on the strength of the two mechanisms.
The advantage scales with the boost, so the boost must be nonzero.
The collision fee must also exceed a threshold set by the boost, the single-trade payoffs, and the composition of opportunities; below that threshold a collision-accepting classical strategy becomes optimal.
For example, Sec.~\ref{sec:repricing-details} measures reference scales for two such contributions, the collision cost from competition for liquidity and the synergy from cross-stock repricing.
The fee and boost used here lie well above these scales, and our modification raises the frequency of joint opportunities well above that of the original record; since both the existence and the size of the gap depend on how often the interactions occur as well as on their strength, the combination was chosen to make the gap clearly visible rather than to reproduce the market's own response.
The reported cumulative payoff combines the simulated trading profit with these imposed interaction terms.

\section{Coordinated backtest implementation}
\label{app:coordinated-backtest}

The NautilusTrader software package builds a simulation from components that interact with the engine, and each other, through a shared message bus.
The basic component is the \code{Actor}, which subscribes to data, such as the quote feeds of each venue, and reacts to each update through event-handling callbacks.
The \code{Strategy} class extends \code{Actor} with order management, so a strategy can additionally submit and cancel orders and track the resulting positions.
The message bus (\code{msgbus}) is the publish--subscribe channel through which all components communicate: the engine delivers market data to subscribers over it, and components may also publish and subscribe to topics of their own.
This last feature is what our coordination layer uses.

To implement two-player coordinated strategies in a backtest, we define a subclass of NautilusTrader's \code{Actor} class, which we refer to as a \code{CoordinatingActor}, as well as a subclass of \code{Strategy}, the \code{CoordinatedStrategy}.
Together these form the base architecture for implementing coordinated strategies in the backtesting environment.
The \code{CoordinatedStrategy} and \code{CoordinatingActor} instances communicate via NautilusTrader's \code{msgbus}, as depicted in Fig.~\ref{fig:coined-backtest}.
When an agent detects an opportunity, its strategy publishes a coordination request carrying its local observation; the actor responds with a coordination signal tagged with the requesting agent's name, and each strategy bases its decision only on the signal addressed to it.

Each agent runs one instance of \code{CoordinatedStrategy} per asset, which scans the quote ticks of the agent's two venues for price dislocations.
When an arbitrage opportunity is identified the agent requests a signal from the \code{CoordinatingActor} before acting, so the actor gates the strategy's behavior.
Classical coordination is implemented by the \code{ClaimSetActor}, which assigns each agent a fixed set of symbols on which it may act; when the two sets are disjoint, forming a partition of the symbols, the agents never act on the same asset.
Shared randomness need not be simulated directly: a claim-set drawn at random once per run is a mixture of deterministic runs, so its expected payoff is the corresponding weighted average of their payoffs, and the deterministic assignments suffice to evaluate it and to identify the best classical strategy.

To access quantum correlations, we equip the \code{CoordinatingActor} with a two-qubit state, one qubit per agent, and call the resulting class the \code{SpookyActor}.
Each coordination window begins from a fresh copy of the shared state.
When an agent requests a signal for an asset, the actor performs a projective measurement on that agent's qubit in the basis assigned to the asset, samples the outcome by the Born rule, replaces the joint state with its post-measurement state, and reports the outcome to the agent: $1$ for act and $0$ for yield.
A later request from the partner in the same window is measured against this conditioned state, which reproduces the correlations of local measurements on a shared entangled pair.
Each agent's qubit is measured at most once per window, preserving the one-input, one-outcome structure of the game.
A further request in the same window for the same asset replays the recorded outcome, while one for a different asset is answered with $0$ rather than triggering a second measurement in another basis.

\section{Correlated dynamics: measurement details}
\label{sec:repricing-details}

Tables~\ref{tab:collision-penalty} and~\ref{tab:mu-ceiling} provide supporting data for the measurements in Sec.~\ref{sec:realistic-games} of the main text. For repricing, we select isolated aggressive trades in the initiating stock at \code{ARCX}, with no other trade in that stock within $50\,\mathrm{ms}$ and a flat book in the responding stock for the preceding $50\,\mathrm{ms}$. Observations are grouped by directed stock pair, the responding stock's pre-event spread, and response horizon $\tau$, requiring at least fifty moved books per cell. The control applies the same procedure to quote moves in the initiating stock without a trade. Table~\ref{tab:mu-ceiling} reports $\tau=100\,\mu\mathrm{s}$ and spreads of at least two ticks. This response horizon is distinct from feed latency $L$, the collision-counting window $w_{\mathrm{col}}$, and synchronization width $w$.

Let $\Delta m$ be the midpoint shift in the responding stock along the initiating trade's direction and $\Delta s$ its spread change. The opposite- and same-direction payoff changes are $\Delta m-\Delta s/2$ and $-\Delta m-\Delta s/2$, respectively, expressed in ticks of the responding stock. The table reports the mean response in each sample and the take-minus-control difference; the opposite-direction difference is the optimistic synergy estimate discussed in the main text.
Reported means pool all qualifying events across the twenty trading days, including zero-move events, which contribute zero. Days with more qualifying events therefore receive greater weight. Each day-bootstrap replicate resamples trading days and recomputes the same pooled mean.

Uncertainties use 4000 bootstrap replicates of the twenty trading days, with the same resampled days for the take and control samples. Their difference is calculated within each replicate; reported standard errors include a factor of $\sqrt{20/19}$, and differences are computed before rounding. Displayed depth is the median size at the improved quote over improved events in the take sample. Multiplying mean repricing by this depth gives only an indicative monetary scale, since executable profit depends on their event-level association and execution frictions.

\begin{table}[t]
\centering
\caption{Signed payoff change from liquidity depletion versus feed latency over the twenty trading days. 
Base is the non-interacting joint payoff per collision. 
The payoff-change row reports the daily mean with standard error over days; negative values indicate a collision cost. 
Brackets express the change as a fraction of the base. 
The control is the same difference on windows in which only one agent acted.}
\label{tab:collision-penalty}
\footnotesize
\setlength{\tabcolsep}{3pt}
\resizebox{\columnwidth}{!}{%
\begin{tabular}{lcccc}
\hline\hline
 & $L=0$ & $100\,\mu$s & $375\,\mu$s & $1$ ms \\
\hline
collisions (20 days)          & 383 & 388 & 389 & 388 \\
base (\$/collision)           & $+18.3$ & $+4.2$ & $-0.2$ & $-4.9$ \\
payoff change (\$/collision)            & $-0.8\pm0.5$ & $-2.6\pm1.1$ & $-0.7\pm0.7$ & $+0.5\pm0.4$ \\
\quad relative to base        & $[-4\%]$ & $[-60\%]$ & -- & -- \\
control (\$/window)           & $0.00$ & $-0.02$ & $+0.03$ & $+0.05$ \\
\hline\hline
\end{tabular}}
\end{table}

\begin{table*}[t]
\centering
\caption{Second-leg payoff change at $\tau=100\,\mu\mathrm{s}$ with the responding stock's
book at a spread of two or more ticks, in ticks of that stock, with positive values favorable to
the leg, after an isolated aggressive trade in the initiating stock (take) and after a quote
move in that stock with no trade (control). Stock pairs are listed as initiating $\to$ responding stock. Parentheses give one standard error in the
last digits; all three in each leg come from the same bootstrap over trading
days described in the text, and differences are computed before rounding. The
opposite-leg difference is the optimistic estimate of potential synergy under
the attribution and execution assumptions described in the text; it is a
conditional comparison, not a causal effect. Depth is the median displayed
size at the improved touch, over improved events in the take sample.}
\label{tab:mu-ceiling}
\footnotesize
\setlength{\tabcolsep}{4pt}
\begin{tabular}{lccc@{\hspace{10pt}}cccc}
\hline\hline
 & \multicolumn{3}{c}{opposite-direction leg} & \multicolumn{3}{c}{same-direction leg} & \\
\cline{2-4}\cline{5-7}
Stock pair & take & control & take$-$control
    & take & control & take$-$control
    & depth (shares) \\
\hline
AMD$\to$INTC & $+0.0112(20)$ & $+0.0046(9)$ & $+0.0066(16)$
    & $-0.0007(4)$ & $+0.0004(2)$ & $-0.0011(3)$
    & 300 \\
AMD$\to$MU & $+0.0101(13)$ & $+0.0039(4)$ & $+0.0062(10)$
    & $-0.0102(10)$ & $-0.0035(3)$ & $-0.0067(10)$
    & 100 \\
INTC$\to$AMD & $+0.0090(10)$ & $+0.0049(11)$ & $+0.0041(12)$
    & $-0.0118(14)$ & $-0.0043(8)$ & $-0.0075(13)$
    & 100 \\
INTC$\to$MU & $+0.0105(21)$ & $+0.0038(5)$ & $+0.0067(18)$
    & $-0.0095(11)$ & $-0.0041(9)$ & $-0.0054(13)$
    & 90 \\
MU$\to$AMD & $+0.0131(16)$ & $+0.0049(4)$ & $+0.0083(15)$
    & $-0.0154(19)$ & $-0.0048(5)$ & $-0.0106(16)$
    & 100 \\
MU$\to$INTC & $+0.0066(14)$ & $+0.0042(6)$ & $+0.0024(12)$
    & $-0.0007(8)$ & $+0.0004(2)$ & $-0.0012(8)$
    & 200 \\
\hline\hline
\end{tabular}
\end{table*}

\section{Semidefinite programming formulation}
\label{app:sdp}

To compute rigorous upper bounds on the quantum payoff via the NPA hierarchy, we must first express the agents' quantum strategies in a standard algebraic form. 
Naimark's dilation theorem guarantees that any general measurement strategy can be mathematically modeled using simpler projective measurements, provided we allow the shared entangled state to reside in a Hilbert space of unrestricted dimension \cite{Naimark1943povm}. 
Consequently, we can represent the agents' general measurements using sets of orthogonal projectors. 
This projective mapping forms the foundation of our algebraic setup, enabling us to systematically eliminate redundant outcomes via completeness~\cite{CollinsGisin2004} and construct the moment relaxation below.

\textbf{Moment relaxation.}
For the two-agent games, write $x,y$ for the inputs and $a,b$ for the outputs. 
The objective is the payoff $\Pi[p]$ in Eq.~\eqref{eq:classical-obj}, with coefficients $P(x,y)V(a,b,x,y)$. 
Local POVMs can be dilated to projective measurements on enlarged local spaces~\cite{Naimark1943povm}. 
Introduce Alice's projectors $E_a^x$ and Bob's projectors $F_b^y$, suppressing tensor-product identities. 
One outcome per measurement may be eliminated through completeness~\cite{CollinsGisin2004}:
\begin{equation}
 E_{k_A-1}^x=I-\sum_{a<k_A-1}E_a^x,\qquad
 F_{k_B-1}^y=I-\sum_{b<k_B-1}F_b^y.
\end{equation}
The remaining generators are Hermitian and satisfy
\begin{equation}
\begin{gathered}
 E_a^xE_{a'}^x=\delta_{aa'}E_a^x,\quad
 F_b^yF_{b'}^y=\delta_{bb'}F_b^y,\\
 [E_a^x,F_b^y]=0.
\end{gathered}
\label{eq:npa-relations}
\end{equation}
For a finite set $\mathcal S$ of words containing $I$, define
$\Gamma_{u,v}=\langle\psi|u^\dagger v|\psi\rangle$.
Every quantum realization gives $\Gamma\succeq0$ and $\Gamma_{I,I}=1$. 
Products identified by Eq.~\eqref{eq:npa-relations} have equal moments, and products reducing to zero have zero moments. 
For retained outcomes, $p_\Gamma(a,b\mid x,y)=\Gamma_{E_a^x,F_b^y}$; all other probabilities follow by completeness. 
For example,
\begin{equation}
 p_\Gamma(a,k_B-1\mid x,y)=\Gamma_{I,E_a^x}
 -\sum_{b<k_B-1}\Gamma_{E_a^x,F_b^y}.
\end{equation}
Writing $\mathcal L(\Gamma)=0$ for these linear relations, the relaxation is
\begin{equation}
\begin{aligned}
 Q_{\mathrm{ub}}^{(\mathcal S)}=\max_\Gamma\;&
 \sum_{a,b,x,y}P(x,y)V(a,b,x,y)p_\Gamma(a,b\mid x,y)\\
 \text{subject to }&\Gamma\succeq0,\quad\Gamma_{I,I}=1,\quad
 \mathcal L(\Gamma)=0.
\end{aligned}
\label{eq:npa-sdp}
\end{equation}
This exact SDP optimum upper-bounds the tensor-product quantum value $Q$. 
A real symmetric matrix suffices for this real objective and these real constraints: averaging any feasible Hermitian matrix with its complex conjugate preserves feasibility and the objective~\cite{Navascues2008}. 

\textbf{Word sets.}
Let $\mathcal E=\{E_a^x\}$ and $\mathcal F=\{F_b^y\}$ denote the retained generators. 
Products below range over all indicated generators and are reduced using Eq.~\eqref{eq:npa-relations}:
\begin{align}
 \mathcal S_{\mathrm{L1}}&=\{I\}\cup\mathcal E\cup\mathcal F,\nonumber\\
 \mathcal S_{\mathrm{L1+AB}}&=\mathcal S_{\mathrm{L1}}\cup\mathcal E\mathcal F,\nonumber\\
 \mathcal S_{\mathrm{L2}}&=\mathcal S_{\mathrm{L1+AB}}
 \cup\mathcal E\mathcal E\cup\mathcal F\mathcal F,\nonumber\\
 \mathcal S_{\mathrm{L2+AB}}&=\mathcal S_{\mathrm{L2}}\cup\mathcal T_3,\quad
 \mathcal T_3\subseteq\mathcal E\mathcal E\mathcal F\cup\mathcal E\mathcal F\mathcal F.
\label{eq:npa-levels}
\end{align}
Here L2$+$AB denotes the extension by selected length-three words. 
For nested sets, the exact optima obey Eq.~\eqref{eq:npa-ordering-results}. 
Adding only $EF$ words to L2 changes nothing, since those words are already present.

At bare L1, joint moments occur off the diagonal and need not be nonnegative unless positivity constraints are added explicitly. 
At L1$+$AB, $\Gamma_{EF,EF}=\langle EF\rangle\geq0$. 
Completeness supplies the same conclusion for the eliminated outcomes, since their product vectors are linear combinations of the indexed words. 
Thus L1$+$AB yields normalized, nonnegative behaviors, although these need not be quantum realizable.

\textbf{No-opportunity inputs.}
The game action rule is $a=0$ when $x=0$, and $b=0$ when $y=0$. 
It can be enforced directly by fixing $M_1^0=N_1^0=0$ and $M_0^0=N_0^0=I$. 
Alternatively, one may retain a redundant action label at input zero and define the effective action
$\bar a=a\mathbf1_{x\ne0}$, $\bar b=b\mathbf1_{y\ne0}$, using
\begin{equation}
 \widetilde V(a,b,x,y)=V(\bar a,\bar b,x,y)
\end{equation}
to preserve the value of the restricted game. Setting the redundant output to zero leaves every payoff unchanged for both classical and quantum strategies. 

\textbf{Numerical values and precision.}
\code{ncpol2sdpa}~1.14.0~\cite{Wittek2015ncpol}, \code{CVXPY}~1.9.2~\cite{Diamond2016cvxpy}, and \code{Clarabel}~0.11.1~\cite{Goulart2024clarabel} are used to solve the SDP with feasibility and duality-gap tolerances of $10^{-7}$. 
For a nonzero objective, positive scaling by $s=\max_{a,b,x,y}|P(x,y)V(a,b,x,y)|$ preserves the optimizer; the returned objective must be multiplied by $s$. 
If $s=0$, the objective is identically zero and no scaling is needed.

\subsection{Comparison of finite NPA levels}
\label{sec:npa-convergence}

\begin{figure*}[t]
\centering
\includegraphics[width=\textwidth]{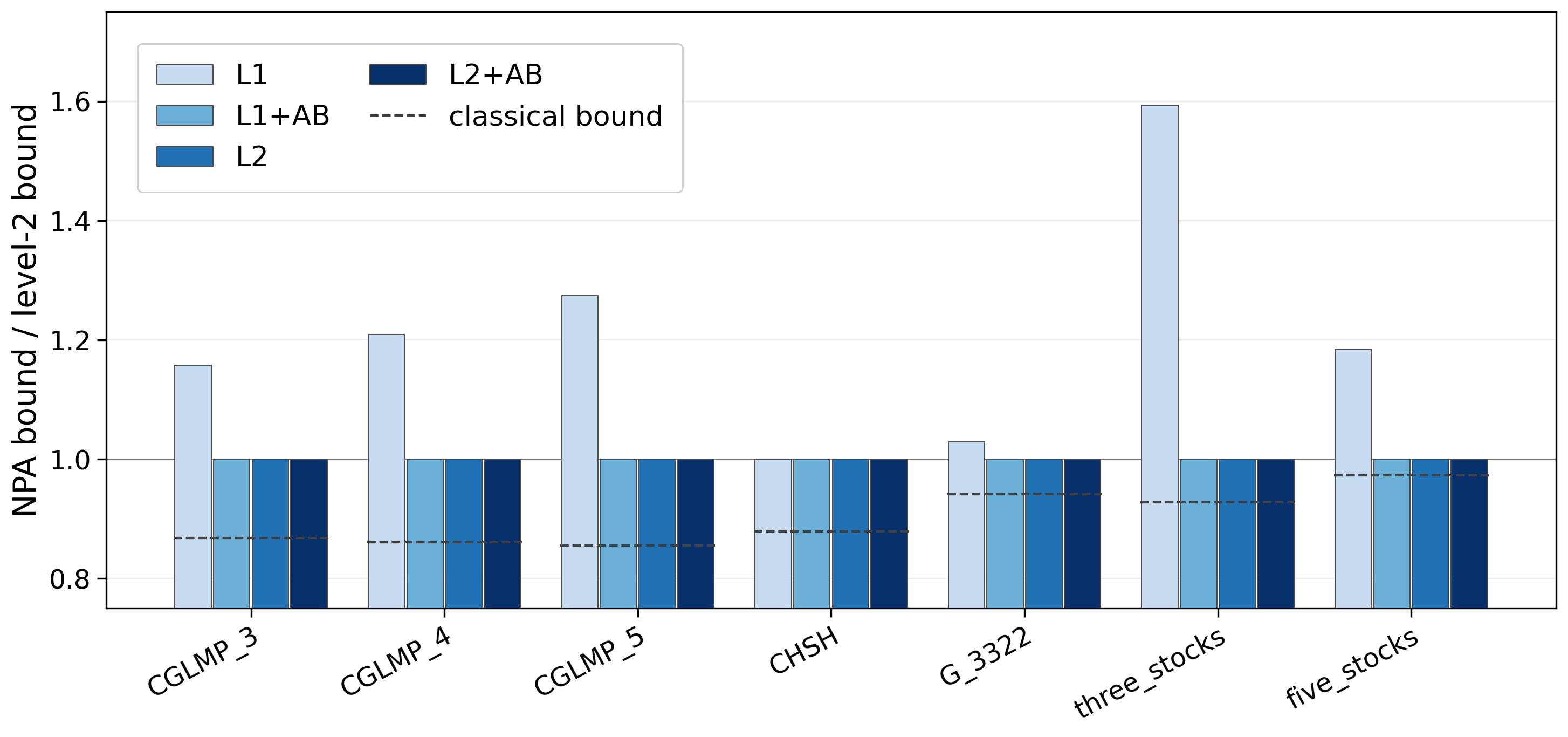}
\caption{Reported NPA estimates normalized by the L2 estimate; dashed segments show the similarly normalized classical optimum. 
The labels three\_stocks and five\_stocks denote instances of the Brandenburger--La Mura (BLaM) game~\cite{blm2015} with three and five stocks, respectively.} 
\label{fig:npa-validation-convergence}
\end{figure*}

Figure~\ref{fig:npa-validation-convergence} compares the NPA estimates across L1, L1$+$AB, L2, and L2$+$AB for seven benchmark games (CHSH, CGLMP$_3$--CGLMP$_5$, $G_{3322}$, and the three- and five-stock games). The certified upper bounds exceed raw solver estimates by at most $10^{-5}$ relative. Except for $G_{3322}$ (which lies $1.25\times10^{-4}$ above L2), L1$+$AB matches L2 to within $8.4\times10^{-7}$ relative, motivating L1$+$AB for screening and L2 for tighter checks.

All Bell-game estimates align with known reference values (CHSH matches $(2+\sqrt2)/4$; CGLMP$_3$ rounds to $0.8644$~\cite{Araujo2020}; $G_{3322}$ L2 rounds to $0.8502$). For CHSH and CGLMP$_3$--CGLMP$_5$, the L2 bounds match our best QSS payoffs (Sec.~\ref{app:see-saw}) to within $1.3\times10^{-7}$ relative, consistent with the true quantum value to the reported numerical precision.

\section{The Quantum Strategy Search algorithm implementation}
\label{app:see-saw}
The QSS algorithm alternates conditional maximizations over the state and the two agents' measurements of the payoff at a fixed local dimension $d$. 
One complete sweep consists of a state update, Alice's measurement update, a state update, Bob's measurement update, and a final state update.

\textbf{Core algorithm updates.}
For fixed measurements, the state is updated to the top eigenvector of $\Omega$ [Eq.~\eqref{eq:qss-payoff-operator}]. 
The measurement updates depend on the number of available outcomes. 
For binary outcomes, the update is simply the projector onto the positive eigenspace of $G_x^0 - G_x^1$. 
For multi-outcome scenarios, the update is framed as a semidefinite program (SDP) per input and solved using Clarabel. 
To ensure every iterate remains an exact POVM, the SDP solution is mapped back onto the POVM set by clipping the negative eigenvalues of each effect and applying $S^{-1/2}(\cdot)S^{-1/2}$ with $S = \sum_a M_a$.

\textbf{Analytical Measurement Updates for Two-Action Regimes.} 
While the local measurement updates generally require solving a semidefinite program (SDP), an efficient analytical shortcut exists when an agent is restricted to two available actions (labeled $0$ and $1$). Crucially, this positive-eigenspace update applies to games with any number of agents.
Under these binary-action conditions, the optimal measurement update for a given input $x_i$ avoids a full SDP. Instead, the optimal POVM element $M_0^{x_i}$ is simply the projector onto the positive eigenspace of the operator difference $G_0^{x_i} - G_1^{x_i}$, while the second element is strictly determined by completeness: $M_1^{x_i} = I_i - M_0^{x_i}$. Zero-eigenvalue directions can be arbitrarily assigned to either outcome without altering the total expected payoff~\cite{LiangDoherty2007}.

\textbf{Initialization and restarts.} 
Each search runs $R$ independent restarts with a fixed seed. 
The initial search (restart zero) is seeded precisely to the optimal classical strategy. 
Classical seeds serve only to guarantee the classical floor (Theorem~\ref{thm:qso-classical-floor}), whereas all quantum advantages discovered by QSS are driven entirely by random restarts. In all numerical experiments across all twelve benchmark games, searches seeded exclusively at the classical strategy remained trapped at $C$ throughout all subsequent sweeps.
Subsequent restarts draw a Haar-random unitary for each input of each party, assigning its columns to the outcomes in a balanced way so that every outcome receives at least one basis vector whenever $d \ge k$. 
This balanced assignment is structurally important: an outcome that receives the whole space constitutes a deterministic response, which is stationary. 
Regardless of the seed, the initial state is not drawn randomly; the first sweep always begins directly with the exact state update for the initialized measurements.

\begin{theorem}[Classical Floor Guarantee]
\label{thm:qso-classical-floor}
For a finite game with classical optimum $C$, prescribe $d\geq1$ and allow all local POVMs consistent with the game's local action restrictions. 
If the candidate set contains a classical maximizing response pair and the candidate with greatest feasible payoff is returned, then $\Pi[p_{\mathrm{retained}}]\geq C$.
\end{theorem}
\begin{proof}
Let $(f_\star,g_\star)$ attain Eq.~\eqref{eq:classical-det}, and set $M_a^x=\delta_{a,f_\star(x)}I_d$ and $N_b^y=\delta_{b,g_\star(y)}I_d$.
These effects are positive semidefinite, sum to the identity for each input, and obey the action restrictions because $f_\star$ and $g_\star$ select permitted actions. 
For any normalized $|\psi\rangle\in\mathbb C^d\otimes\mathbb C^d$, Eq.~\eqref{eq:qso-born} gives $p(a,b\mid x,y)=\delta_{a,f_\star(x)}\delta_{b,g_\star(y)}\langle\psi|\psi\rangle =\delta_{a,f_\star(x)}\delta_{b,g_\star(y)}$, so the payoff is $\sum_{x,y}P(x,y)V(f_\star(x),g_\star(y),x,y)=C$. 
Selecting the greatest payoff from a set that contains this candidate gives $\Pi[p_{\mathrm{retained}}]\geq C$.
\end{proof}

\textbf{Exact and numerical updates.} The state update and the two-outcome measurement update are closed-form eigendecompositions, exact up to floating-point rounding, so the payoff cannot decrease. 
For more than two outcomes, the measurement update is a numerical SDP solved to $10^{-9}$. 
We make its solution a valid measurement by removing small negative eigenvalues and renormalizing, which can slightly change the payoff. 
The update is therefore kept only if it does not lower the payoff, so monotonicity on this path holds only to numerical precision. 
Every quantum improvement reported here comes from a random restart. 
The classical seed only guarantees that the returned payoff is not below $C$, to the same precision.

\textbf{Execution parameters and termination.} 
The gap scan of Fig.~\ref{fig:doped-gap-scan} uses $R=12$ restarts at local dimension $d=2$.

\section{Reconciling differences in the cumulative backtest payouts}
\label{app:backtest-reconciliation}

We account for the differences between the engine results and the predictions of the $(P,V)$ tensor model in three steps.
We rule out the strategy itself as their source, identify the mechanism responsible and how its effect is distributed across the twenty days, and trace its consequence for the measured advantage.

The strategy is not the source.
On every day's realized tensor, the optimized singlet strategy lies within $0.025\%$ of the NPA level-2 bound, measured as a share of the classical-to-NPA gap.
Netted over the twenty days, its P,V prediction is $\$303{,}686.3$, only $\$2.3$ short of the NPA level-2 bound, while its backtest realizes $\$286{,}323.6$, falling $\$17{,}365$ short of the bound.
The choice of strategy therefore accounts for $\$2.3$ of the backtest's $\$17{,}365$ in uncaptured payoff, a fraction of $1.3\times10^{-4}$.
The remaining $\$17{,}363$ shortfall of the backtest below its $(P,V)$ tensor prediction arises in execution.

A measurable contribution to the shortfall is attributable to the strategy's stale-liquidity guard, which prevents repeated trades against liquidity already consumed by an agent.
The backtesting engine does not alter the historical record, so a quote an agent has already traded against remains displayed after the trade; without a guard the agent could trade the same displayed liquidity again and again, collecting profit the market would never have offered.
The guard therefore keeps a record of the liquidity each agent has taken at each price level and withholds further orders against it until the venue's quote changes.
On the modified records this has a side effect.
An injected opportunity is often paired with a real quote that the agent consumed earlier and that has not since refreshed, leading to the guard's suppression of the trade.
Table~\ref{tab:guard-accounting} shows that this mechanism accounts for nearly all of the classical arm's shortfall on the days where it is strongest.

\begin{table*}[t]
\caption{Guard accounting for the four days on which guard-attributed losses exceed $1\%$ of the classical model payoff.
Suppression is the fraction of injected rounds on which the guard withheld an order.
The realized ratio is the engine's realized payoff divided by the model prediction (quantum: mean over 25 seeds, $\pm 2$ s.e.).
For the classical arm, which is deterministic, the shortfall (model minus engine) is decomposed round by round: every round whose intended actions did not execute is attributed to the reason it was silenced, and the stale-counterpart column gives the share of that loss in which the guard blocked the real counterpart quote of an injected opportunity.
The part of the shortfall not attributed to suppressed rounds is under $0.4\%$ of the model payoff on each of these days.}
\label{tab:guard-accounting}
\begin{ruledtabular}
\begin{tabular}{lrrrrrrr}
 & & \multicolumn{2}{c}{suppression} & \multicolumn{2}{c}{realized ratio} & \multicolumn{2}{c}{classical shortfall} \\
\cline{3-4} \cline{5-6} \cline{7-8}
date & rounds & classical & quantum & classical & quantum & (\$) & stale counterpart \\
\hline
2024-08-07 & 733 & 0.5\% & 1.2\% & 0.987 & $0.966 \pm 0.024$ & 100 & 99.7\% \\
2026-03-06 & 1439 & 1.5\% & 2.6\% & 0.960 & $0.960 \pm 0.018$ & 597 & 99.7\% \\
2026-03-27 & 2130 & 3.0\% & 5.6\% & 0.906 & $0.889 \pm 0.014$ & 2096 & 99.8\% \\
2026-03-30 & 5720 & 4.1\% & 6.6\% & 0.776 & $0.872 \pm 0.009$ & 13298 & 92.7\% \\
\end{tabular}
\end{ruledtabular}
\end{table*}

The guard's effect falls into two regimes.
On sixteen days the classical arm realizes its predicted payoff almost exactly ($0.998$ in aggregate), while the quantum arm realizes $0.972$, a shortfall of about $3\%$ that does not depend on how dense the record is; the guard costs the quantum arm more because its advantage sits in joint rounds that require both agents to act.
On the four days of Table~\ref{tab:guard-accounting}, where the guard is strongest, both arms lose more ($0.846$ classical and $0.895$ quantum in aggregate, with the ordering varying by day), so the two population ratios are nearly equal ($0.943$ quantum and $0.940$ classical) only because the regimes offset.

These regimes are what separate the realized advantage from the predicted one.
Since the injection recipe fixes the composition of opportunities, the predicted relative advantage is nearly the same on every day: across the 20 days its mean and standard deviation are a $11.2 \pm 0.6\%$.
The realized advantage, $9.4 \pm 4.3\%$ (likewise across days), varies far more, because it also carries each day's realized ratios, and through them each original record's own market dynamics.
This pattern places the problem in the interaction between the injected events and the guard, rather than in the strategy or the model.
The effect is strongest where injected opportunities are most crowded, so that many of them depend on the same real counterpart quote before that quote refreshes.
A refined injection protocol that accounts for the guard's effect on the counterpart venue should reduce it.

\end{document}